\documentclass[preprint,superscriptaddress,nofootinbib]{revtex4-1}
\usepackage{hyperref}
\usepackage{bbm}
\usepackage{amsfonts}
\usepackage{mathrsfs}
\usepackage{mathtools}
\usepackage{bbold}

\usepackage{amsmath,amssymb}

\usepackage{epsfig}
\usepackage{color}
\usepackage{graphicx}    
\usepackage{url}
\usepackage{float}
\usepackage{pstricks}
\usepackage{multirow}
\usepackage{slashed}

\newcommand{\be}{\begin{equation}}
\newcommand{\ee}{\end{equation}}
\newcommand{\nl}{\nonumber \\}
\newcommand{\x}{\chi}
\newcommand{\xii}{\chi_i}

\newcommand{\xp}{\chi^+}

\newcommand{\xm}{\chi^-}

\newcommand{\order}{\mathcal{O}}

\begin{document}
\title{The Diphoton and Diboson Excesses in a Left-Right Symmetric Theory of Dark Matter}
\author{Asher Berlin}
\affiliation{Department of Physics, Enrico Fermi Institute \\
\vskip -8 pt
University of Chicago, Chicago, IL 60637 USA}

\begin{abstract}
We explore the possibility that the recently reported diphoton excess at ATLAS and CMS can be accommodated within a minimal extension of a left-right symmetric model. Our setup is able to simultaneously explain the Run 2 diphoton and Run 1 diboson excesses, while providing a standard thermal freeze-out of weak-scale dark matter. In this scenario, the 750 GeV neutral right-handed Higgs triplet is responsible for the diphoton excess. Interactions of this state with the neutral and charged components of dark matter multiplets provide the dominant mechanisms for production and decay. A striking signature of this model is the additional presence of missing energy in the diphoton channel.
\end{abstract}

\preprint{}

\maketitle


\section{\textbf{Introduction}}
\label{sec:introduction}

The ATLAS and CMS collaborations both recently reported an excess in the diphoton mass distribution around 750 GeV with 3.2 $\text{fb}^{-1}$ and 2.6 $\text{fb}^{-1}$ of 13 TeV data, respectively~\cite{ATLAS-CONF-2015-081,CMS-PAS-EXO-15-004}. This signal is consistent with a rate of $R_{\gamma \gamma} \approx 1-10$ fb, and the local significance quoted by ATLAS and CMS is $3.6 \sigma$ and $2.6 \sigma$, respectively, for the case of a narrow resonance. Under the assumption of a larger width, $\Gamma \sim 45$ GeV, the local significance shifts to $3.9 \sigma$ for ATLAS and $2.0 \sigma$ for CMS. Although it is still plausible that these excess events, or some fraction, correspond to an upward statistical fluctuation or systematic issue, the possibility that this signal constitutes physics beyond the standard model (SM) has resulted in a flurry of recent model building attempts, averaging approximately six new papers per day~\cite{Backovic:2015fnp,Nakai:2015ptz,Buttazzo:2015txu,Franceschini:2015kwy,Knapen:2015dap,Ellis:2015oso,Gupta:2015zzs,Molinaro:2015cwg,McDermott:2015sck,Low:2015qep,Petersson:2015mkr,Dutta:2015wqh,Kobakhidze:2015ldh,Cox:2015ckc,Ahmed:2015uqt,Agrawal:2015dbf,Becirevic:2015fmu,No:2015bsn,Demidov:2015zqn,Chao:2015ttq,Fichet:2015vvy,Curtin:2015jcv,Bian:2015kjt,Csaki:2015vek,Falkowski:2015swt,Bai:2015nbs,Ghosh:2015apa,Kim:2015ron,Alves:2015jgx,Megias:2015ory,Bernon:2015abk,Chao:2015nsm,Ringwald:2015dsf,Arun:2015ubr,Han:2015cty,Chang:2015bzc,Dhuria:2015ufo,Han:2015dlp,Luo:2015yio,Chang:2015sdy,Bardhan:2015hcr,Feng:2015wil,Barducci:2015gtd,Chakraborty:2015jvs,Ding:2015rxx,Hatanaka:2015qjo,Antipin:2015kgh,Wang:2015kuj,Cao:2015twy,Huang:2015evq,Bi:2015uqd,Kim:2015ksf,Berthier:2015vbb,Cline:2015msi,Chala:2015cev,Kulkarni:2015gzu,Dev:2015isx,Murphy:2015kag,Hernandez:2015ywg,Dey:2015bur,Pelaggi:2015knk,Huang:2015rkj,Chabab:2015nel,Cao:2015xjz,Moretti:2015pbj,Patel:2015ulo,Badziak:2015zez,Chakraborty:2015gyj,Cvetic:2015vit,Allanach:2015ixl,Davoudiasl:2015cuo,Das:2015enc,Cheung:2015cug,Liu:2015yec,Zhang:2015uuo,Casas:2015blx,Hall:2015xds,Han:2015yjk,Park:2015ysf,Chway:2015lzg,DelDebbio:2015byq,Salvio:2015jgu,Li:2015jwd,Son:2015vfl,An:2015cgp,Wang:2015omi,Cao:2015scs,Gao:2015igz,Dev:2015vjd,Tang:2015eko,Cao:2015apa,Cai:2015hzc,Kim:2015xyn,Chao:2015nac,Bi:2015lcf,Anchordoqui:2015jxc,Bizot:2015qqo,Hamada:2015skp,Huang:2015svl,Chiang:2015tqz,Kang:2015roj,Kanemura:2015bli,Dong:2015dxw,Low:2015qho,Hernandez:2015hrt,Jiang:2015oms,Kaneta:2015qpf,Dasgupta:2015pbr,Jung:2015etr,Potter:2016psi,Palti:2016kew,Nomura:2016fzs,Ko:2016lai,Danielsson:2016nyy,Chao:2016mtn,Hernandez:2016rbi,Modak:2016ung,Dutta:2016jqn}. 
Although the diversity of models that are able to accommodate the signal is quite extensive, general conclusions can still be made. For instance, in addition to the resonant particle itself, the large signal rate and mass require new physics involved in the decay~\cite{Knapen:2015dap}. Furthermore, aside from some mild tension, the majority of the signal at 13 TeV is roughly consistent with existing bounds from Run 1 diphoton searches if one assumes a $2 \to 2$ production mechanism, i.e., $p p \to X \to \gamma \gamma$~
\cite{Franceschini:2015kwy}. This tension may be ameliorated, for example, if the 750 GeV resonance is produced from the decay of a somewhat heavier state. 

Thus far, Run 2 has proven to be exciting, but more data is warranted in order to clarify the origin of the excess. Evidence of new physics may also still be present in Run 1 data. One such signal that recently gained interest is a diboson resonance at a mass of $\sim 1.9$ TeV at $\sqrt{s}=8$ TeV~\cite{Aad:2015owa}. An explanation for this and several other Run 1 excesses~\cite{Khachatryan:2014dka,CMS-PAS-EXO-14-010,Khachatryan:2015sja,Khachatryan:2014gha}, e.g., in the dijet channel, has been presented within the context of a left-right symmetric model, extending the SM gauge group to $SU(3)_c \times SU(2)_L \times SU(2)_R \times U(1)_{B-L}$~, with a Higgs sector consisting of a bidoublet scalar and $SU(2)_R$ triplet scalar~\cite{Dobrescu:2015qna,Dobrescu:2015yba,Coloma:2015una,Dobrescu:2015jvn}. In describing the diboson signal, these models predict a new charged gauge boson, $W^\prime$, with a mass close to 2 TeV and an additional neutral gauge boson, $Z^\prime$, with a mass around $3-4$ TeV. While no dijet excess at the same mass has been noted at 13 TeV~\cite{ATLAS:2015nsi}, this may simply be the result of insufficient statistics, and only more data will ultimately be able to weigh in on the issue.

The mild tension between the diphoton searches at 8 and 13 TeV may be hinting towards yet another resonance beyond 750 GeV~\cite{Franceschini:2015kwy}. In this paper, we introduce a model where the heavy $W^\prime$ can explain the Run 1 signals through $p p \to W^\prime \to j j \, , \, W Z \, , \, W h $~, while the $Z^\prime$ and neutral component of the right-handed Higgs triplet ($\Delta^0$) play the role of the additional heavy resonance and 750 GeV diphoton resonance, respectively. In particular, the diphoton signal is generated through the cascade $pp \to Z^\prime \to X ~ Y \to  X ~ X ~ \Delta^0 (\to \gamma \gamma)$~, where at the moment $X$ and $Y$ are some  unspecified soft states. Similar production mechanisms have been studied for example in Ref.~\cite{An:2015cgp}.

Dark matter can also easily be incorporated into left-right symmetric models since they generically involve additional stabilizing symmetries and heavy gauge/Higgs bosons, allowing for new portals between the SM and dark matter sectors~\cite{Aulakh:1998nn,Garcia-Cely:2015quu,Heeck:2015qra,Berlin:2016}. One example of this involves adding several new colorless dark matter multiplets, non-trivially charged under $SU(2)_L \times SU(2)_R \times U(1)_{B-L}$~. Interestingly, as we will see in the following sections, the neutral components of these multiplets may make up the cosmological abundance of dark matter today, while the electrically charged components can couple to $\Delta^0$~, facilitating a large loop-induced branching fraction to pairs of photons, crucial for reproducing the diphoton signal. 

The remainder of this paper is structured as follows. Sec.~\ref{sec:model} briefly reviews left-right symmetric models and extends the minimal model with several new dark matter multiplets. In Sec.~\ref{sec:diphoton}, we explore the parameter space that is most relevant for accommodating the diphoton and diboson signals, while simultaneously allowing for the standard freeze-out of dark matter. We summarize our results and implications in Sec.~\ref{sec:conclusions}.

\section{\textbf{The Model}}
\label{sec:model}

Adequate breaking of $SU(2)_L \times SU(2)_R \times U(1)_\text{B$-$L}$ down to $U(1)_\text{em}$ requires an extended Higgs sector. For further review, see Ref.~\cite{Duka:1999uc}. The minimal Higgs content consists of one right-handed complex scalar triplet with quantum numbers $\Delta_R : ({\bf 1},{\bf 3},2)$ and a complex scalar bidoublet $\phi : ({\bf 2},{\bf 2},0)$. 

The Higgs triplet $\Delta_R$ breaks $SU(2)_L \times SU(2)_R \times U(1)_\text{B-L}$ down to the SM gauge group $SU(2)_L \times U(1)_Y$ after acquiring a vacuum expectation value (VEV), $v_R$~. The SM-like VEV of the bidoublet $\phi$~, $v = 174$ GeV, further breaks $SU(2)_L \times U(1)_Y$ down to $U(1)_\text{em}$~. We will work in the alignment and decoupling limits where the light Higgs, $h$~, is SM-like and the additional Higgs bosons of $\phi$ are not present in the low-energy theory. The right-handed triplet is parametrized in unitary gauge as
\begin{align}
\Delta_R &= \begin{pmatrix} 0 & & \Delta^{++} \\ v_R + \frac{1}{\sqrt{2}}~ \Delta^0  &  & 0  \end{pmatrix} 
~,
\end{align}
where we have dropped $\order(m_W / m_{W^\prime})$ mixing with the charged Higgs bosons of $\phi$~. For simplicity, we also assume that the doubly charged Higgs, $\Delta^{++}$~, is decoupled. In matching to the observed rate and mass of the Run 1 diboson excess, we require $v_R \sim 3-4 \text{ TeV}$, $g_R \sim 0.45-0.6$, and $\tan{\beta} \sim 0.5-2$, where  $g_R$ is the $SU(2)_R$ gauge coupling and $\tan{\beta}$ is the ratio of VEVs that appear in the diagonal entries of $\phi$~, analogous to that found in a two-Higgs doublet model~\cite{Dobrescu:2015yba}.

The VEVs of $\Delta_R$ and $\phi$ contribute to the masses of the charged and neutral gauge bosons. From here on out, we will work in the limit that $v_R \gg v$~, or equivalently $m_{W^\prime} \gg m_W$~. At leading order, the masses of the $W^\prime$ and $Z^\prime$ are $m_{W^\prime} \approx g_R~v_R$~ and $m_{Z^\prime}^2 \approx 2 \left( g_R^2 + g_{_\text{B-L}}^2 \right) v_R^2$~, where $g_{_\text{B-L}}$ is the $U(1)_\text{B-L}$ gauge coupling, related to the hypercharge gauge coupling by $g^{\prime -2} = g_R^{-2} + g_{_\text{B-L}}^{-2}$~. 

Previous studies of dark matter in left-right symmetric models have largely focused on either SUSY inspired scenarios~\cite{Aulakh:1998nn} or ``pure states" consisting of additional electroweak multiplets with a single Majorana mass~\cite{Garcia-Cely:2015quu,Heeck:2015qra}. In this paper, we instead explore simple models of ``mixed states," incorporating several multiplets, which mass mix after electroweak symmetry breaking (EWSB) through Yukawa interactions with the Higgs sector~\cite{Berlin:2016}. Contrary to the models involving pure states, models of mixed dark matter allow for tree-level interactions between the lightest neutral state and electroweak bosons, which easily facilities a standard cosmological history for varying ranges of dark matter masses. Interestingly, the Higgs Yukawa interactions also couple $\Delta^0$ to the electrically charged components of the dark matter multiplets. This will be important when we discuss the radiative decay of $\Delta^0$ in Sec.~\ref{subsec:decay}.

\begin{table}[t]
\centering
\begin{tabular}{| c || c | c | c |}
\hline
Field & $SU(2)_L \times SU(2)_R \times U(1)_\text{B-L}$ & Spin \\ \hline \hline
$T_1$ & $({\bf 1},{\bf 3},0)$ & 1/2 \\ \hline
$T_2$ & $({\bf 1},{\bf 3},+2)$ & 1/2 \\ \hline
$T_3$ & $({\bf 1},{\bf 3},-2)$ & 1/2 \\ \hline
\end{tabular}
\caption{Field content of the dark matter sector.}
\label{table:fields}
\end{table}

We now define a simple extension to the left-right model described above, by introducing three $SU(2)_R$ triplet Weyl fermions ($T_{1,2,3}$) with charges as shown in Table~\ref{table:fields}. Note that our field content explictly breaks the charge/parity $L \leftrightarrow R$ symmetry implemented in many left-right models. Although one could imagine the inclusion of additional triplets charged under $SU(2)_L$~, we assume for simplicity that any are sufficiently decoupled from the low-energy spectrum. In order to guarantee electrically neutral dark matter, the fields must be evenly charged under $U(1)_\text{B-L}$~, which also stabilizes the lightest neutral components up to arbitrary order. Furthermore, the fields are given appropriate B-L charge so that they may couple to $\Delta_R$ at tree-level, and a vector pair is needed to cancel anomalies. $T_{1,2,3}$ are then parametrized as
\be
T_1 = \begin{pmatrix} t_1^0/\sqrt{2} & t_1^+ \\ t_1^- & -t_1^0/\sqrt{2} \end{pmatrix} ~,~ T_2 = \begin{pmatrix} t_2^+/\sqrt{2} & t_2^{++} \\ t_2^0 & -t_2^+/\sqrt{2} \end{pmatrix} ~,~ T_3 = \begin{pmatrix} t_3^-/\sqrt{2} & t_3^0 \\ t_3^{--} & -t_3^-/\sqrt{2} \end{pmatrix}
~,
\ee
where the $\pm$ superscripts are labels indicating that these components will make up electrically charged fermions, and the factors of $\sqrt{2}$~ ensure canonical normalization of the kinetic terms, $\mathcal{L} \supset \sum\limits_{i=1,2,3} \text{tr} (T_i^\dagger i \bar{\sigma}^\mu D_\mu T_i)$~. The remaining terms in a general renormalizable Lagrangian are
\be
\label{eq:DarkLag}
-\mathcal{L} \supset \frac{1}{2} M_1 ~ \text{tr}(T_1^2) + M_{23} ~ \text{tr}(T_2 T_3) + \lambda_1 ~ \text{tr}(T_3 T_1 \Delta_R) + \lambda_2 ~ \text{tr}(T_2 T_1 \Delta_R^\dagger) + \text{h.c.}
~,
\ee
where 2-component Weyl spinor indices are implied, and traces refer to sums over $SU(2)_R$ indices. $M_1$ and $M_{23}$ are bare triplet mass terms, and $\lambda_{1,2}$ are real dimensionless Yukawa couplings. After EWSB, the neutral and charged mass matrices for the components of $T_{1,2,3}$ are
\begin{align}
\label{eq:massmatrix}
-\mathcal{L} &\supset \frac{1}{2} \begin{pmatrix} t_1^0 & t_2^0 & t_3^0 \end{pmatrix} \begin{pmatrix} M_1 & \lambda_2 v_R / \sqrt{2} & - \lambda_1 v_R / \sqrt{2} \\\lambda_2 v_R / \sqrt{2} & 0 & M_{23} \\ - \lambda_1 v_R / \sqrt{2} & M_{23} & 0 \end{pmatrix} \begin{pmatrix} t_1^0 \\ t_2^0 \\ t_3^0 \end{pmatrix}
\nl
& + \begin{pmatrix} t_1^+ & t_2^+ \end{pmatrix} \begin{pmatrix} M_1 & \lambda_1 v_R / \sqrt{2} \\ -\lambda_2 v_R \sqrt{2} & M_{23} \end{pmatrix} \begin{pmatrix} t_1^- \\ t_3^- \end{pmatrix} + M_{23} ~ t_2^{++} t_3^{--} + \text{h.c.}
~.
\end{align}
The neutral components $t_{1,2,3}^0$ mass mix to give 3 Majorana fermions ($\x_1 \, , \, \x_2 \, , \, \x_3$), the lightest of which ($\x_1$) will be the cosmologically stable dark matter. The singly charged components mix to give two Dirac fermions ($\chi_1^\pm \, , \, \x_2^\pm$), and the doubly charged components enter as a single Dirac fermion of mass $M_{23}$~. To satisfy limits from chargino searches at LEP, we only consider values of $M_{23} > 100$ GeV~\cite{LEP:Chargino}. 

Diagonalizing the neutral mass matrix yields the projection
\be
\label{eq:mixingangle}
\xii = N_{t_1}^i ~ t_1^0 + N_{t_2}^i ~ t_2^0 + N_{t_3}^i ~ t_3^0 \quad (i = 1,2,3)
~,
\ee
and the charged gauge eigenstates are similarly decomposed as
\begin{align}
t_1^- &= U_{11} ~ \xm_1 + U_{12} ~ \xm_2
\nl
t_3^- &= U_{21} ~ \xm_1 + U_{22} ~ \xm_2
\nl
t_1^+ &= V_{11} ~ \xp_1 + V_{12} ~ \xp_2
\nl
t_2^+ &= V_{21} ~ \xp_1 + V_{22} ~ \xp_2
~.
\end{align}
$U_{ij}$ and $V_{ij}$ are orthogonal matrices that are constructed from the eigenvectors of ${\bf M}^\dagger \,  {\bf M}$ and ${\bf M} \,  {\bf M}^\dagger$, respectively, where ${\bf M}$ is the charged mass matrix of Eq.~(\ref{eq:massmatrix}).

\section{\textbf{The Diboson and Diphoton Signals}}
\label{sec:diphoton}

As discussed briefly in Secs.~\ref{sec:introduction} and \ref{sec:model}, several excesses in Run 1 ATLAS~\cite{Aad:2015owa} and CMS~\cite{Khachatryan:2014dka,CMS-PAS-EXO-14-010,Khachatryan:2015sja,Khachatryan:2014gha} data point towards the presence of a $W^\prime$ boson with mass $1.8-2$ TeV~\cite{Dobrescu:2015qna,Dobrescu:2015yba,Coloma:2015una,Dobrescu:2015jvn}. Matching to the observed excess in the CMS dijet~\cite{Khachatryan:2015sja} distribution through the process $p p \to W^\prime \to jj$ requires $g_R \sim 0.45-0.6$, and consistency with the ATLAS diboson signal~\cite{Aad:2015owa} through $pp \to W^\prime \to W Z$ implies $\tan{\beta} \sim 0.5-2$~\cite{Dobrescu:2015yba}. Restricting to this set of values, the left-right model outlined in the previous section is able to accomodate the diboson and dijet events in the Run 1 data. We also note that the introduction of dark matter multiplets allows for new decay modes of the $W^\prime$, which in principle could alter the preferred values of $g_R$ and $\tan{\beta}$ when matching to the observed excess. However, as we will see in Sec.~\ref{sec:prod}, this will not be of concern for the most viable parameter space since it involves dark matter masses $m_{\x_1} > m_{W^\prime} / 2$~. 

\subsection{Production}
\label{sec:prod}

We now proceed in exploring the possibility that the loop-induced decay of a 750 GeV $\Delta^0$ can account for the diphoton excess. Dimensionless couplings, $\alpha$~, in the left-right scalar potential can induce a small mixing between $\Delta^0$ and the SM Higgs $h$~, $\theta \sim \alpha ~ v / v_R$~, potentially leading to $\Delta^0$ production through SM gluon fusion~\cite{Dasgupta:2015pbr}. $\Delta^0$ may also decay directly to top quarks in this scenario, and hence the signal is suppressed by $\Gamma (\Delta^0 \to t \bar{t}\, ) / \Gamma (\Delta^0 \to g g) \sim 10^{-3}$~. As a result, a very large partial width to photons ($\sim 1$ GeV) is needed to generate a sufficient diphoton rate~\cite{Knapen:2015dap}. For this reason, we assume that any mixing induced by $\alpha$ is negligible in regards to the production or decay of $\Delta^0$~. 

Alternatively, one might consider the production of $\Delta^0$ through the cascade $pp \to W^\prime \to W ~ \Delta^0 (\to \gamma \gamma)$~~\cite{Knapen:2015dap}. To leading order in $m_{W} / m_{W^\prime}$~, the relevant term in the Lagrangian takes the form
\be
\mathcal{L} \supset \frac{\sqrt{2} ~ g_R^2 s_{2 \beta}}{g_L} \frac{m_W^2}{m_{W^\prime}} ~ \Delta^0 ~ W^{+ \mu} W_\mu^{\prime -} + \text{h.c.}
~.
\ee
We find that for $m_{W^\prime} \sim \order(1)$ TeV, the $m_{W} / m_{W^\prime}$ suppression of this interaction results in a very small branching fraction, $\text{BR} \left( W^\prime \to W \Delta^0 \right) \sim \order (10^{-5})$~, much too weak to generate a sufficient diphoton rate, even for $\text{BR} \left( \Delta^0 \to \gamma \gamma \right) \sim 1$~. Note that this is unlike the decay $W^\prime \to W h$~, which can occur at the level of a few percent through the unsuppressed term 
\be
- \mathcal{L} \supset g_R m_W s_{2 \beta} ~ h ~ W^{+ \mu} W_\mu^{\prime -} + \text{h.c.}
~.
\ee

\begin{figure}[t]
\begin{center}
\includegraphics[width=0.45\textwidth]{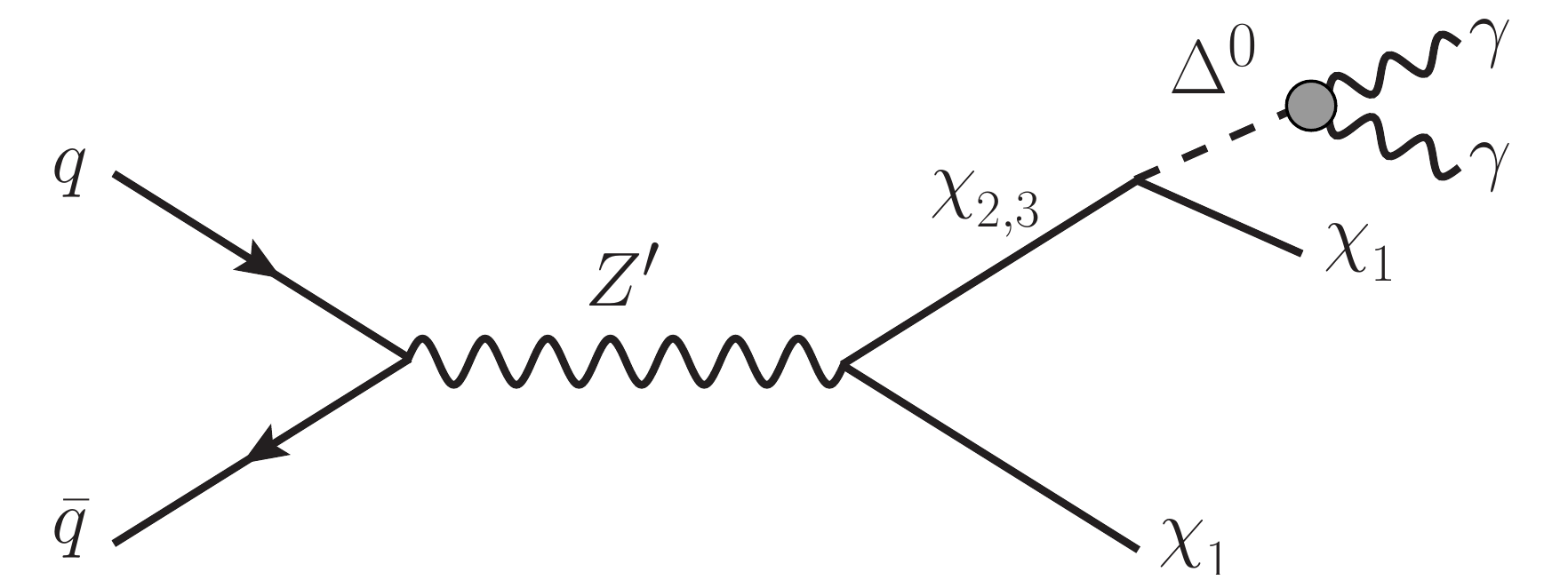} 
\caption{\label{fig:Diagram} Production mechanism for the diphoton signal. The grey circle corresponds to a loop of charged mediators, as discussed in Sec.~\ref{subsec:decay}.}
\end{center}
\end{figure}

As we will see below, $\x_1$ typically freezes out with the proper relic abundance at a mass of $\order(1)$ TeV when $m_{W^\prime} = 1.9$ TeV and $m_{\Delta^0} = 750$ GeV. Therefore, it is natural to imagine that $Z^\prime$ possesses a considerable branching fraction into dark sector states at the level of tens of percent. In light of this insight, we choose to focus on the production of $\Delta^0$ through the cascade $pp \to Z^\prime \to \x_1 ~ \x_{2,3} \to \x_1 ~ \x_1 ~ \Delta^0 (\to \gamma \gamma)$~, as shown in Fig.~\ref{fig:Diagram}. This signal possesses activity beyond the minimal diphoton production in the form of missing energy (MET), and hence the final state $\x_1$'s and $\Delta^0$ should not be significantly boosted. A detailed analysis quantifying the specific degree to which such extra activity can be present in the diphoton channel is beyond the scope of this paper. For simplicity, we assume that the extra MET is softened, and therefore that the mass splittings,
\be
\label{eq:DeltaM}
\Delta M_1 \equiv m_{Z^\prime} - (m_{\x_1} + m_{\x_2}) ~,~ \Delta M_2 \equiv m_{\x_2} - (m_{\x_1} + \Delta^0)
~,
\ee
do not exceed several hundred GeV. 

\begin{figure}[t]
\begin{center}
\includegraphics[width=0.38\textwidth]{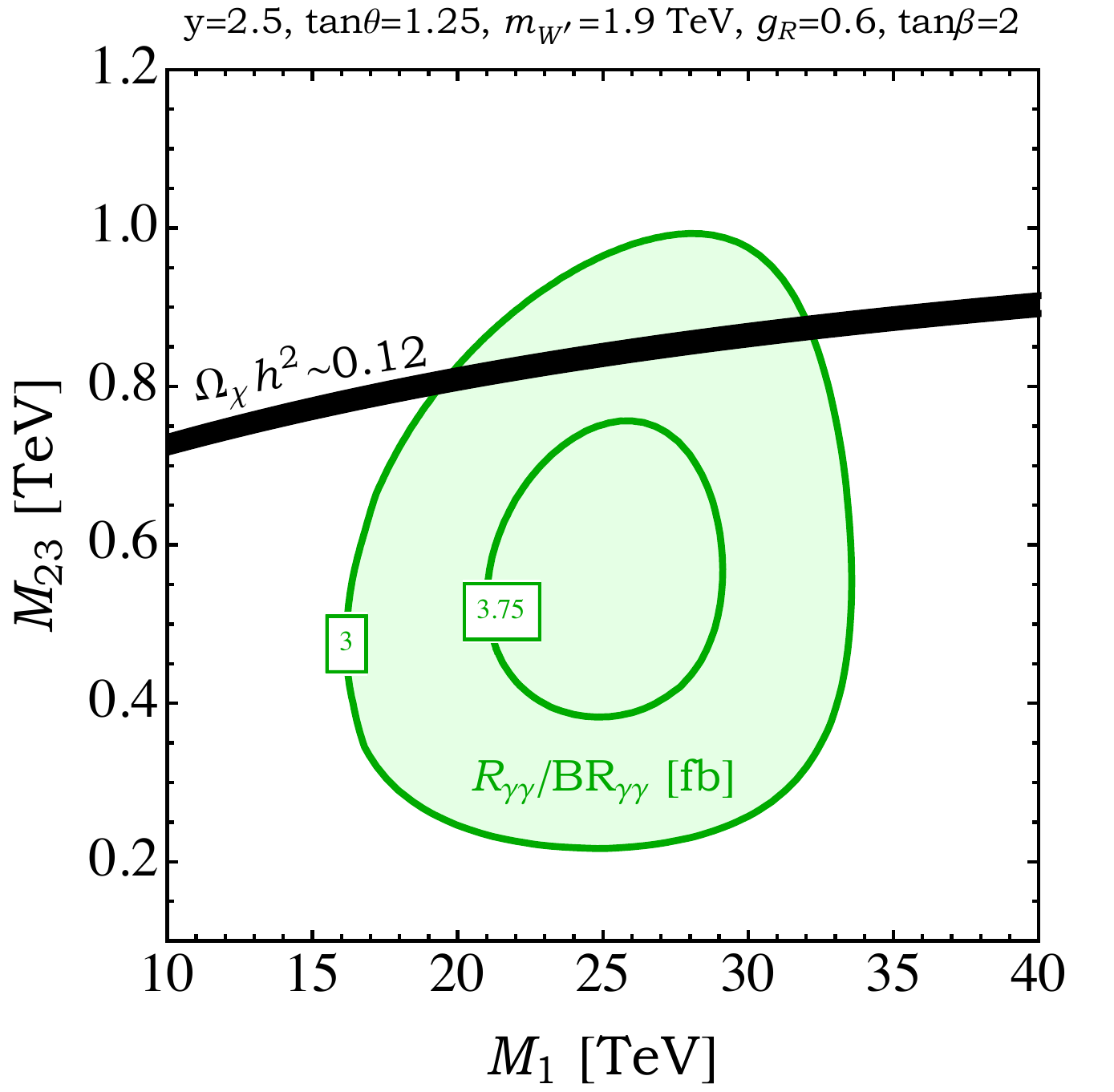}
\includegraphics[width=0.38\textwidth]{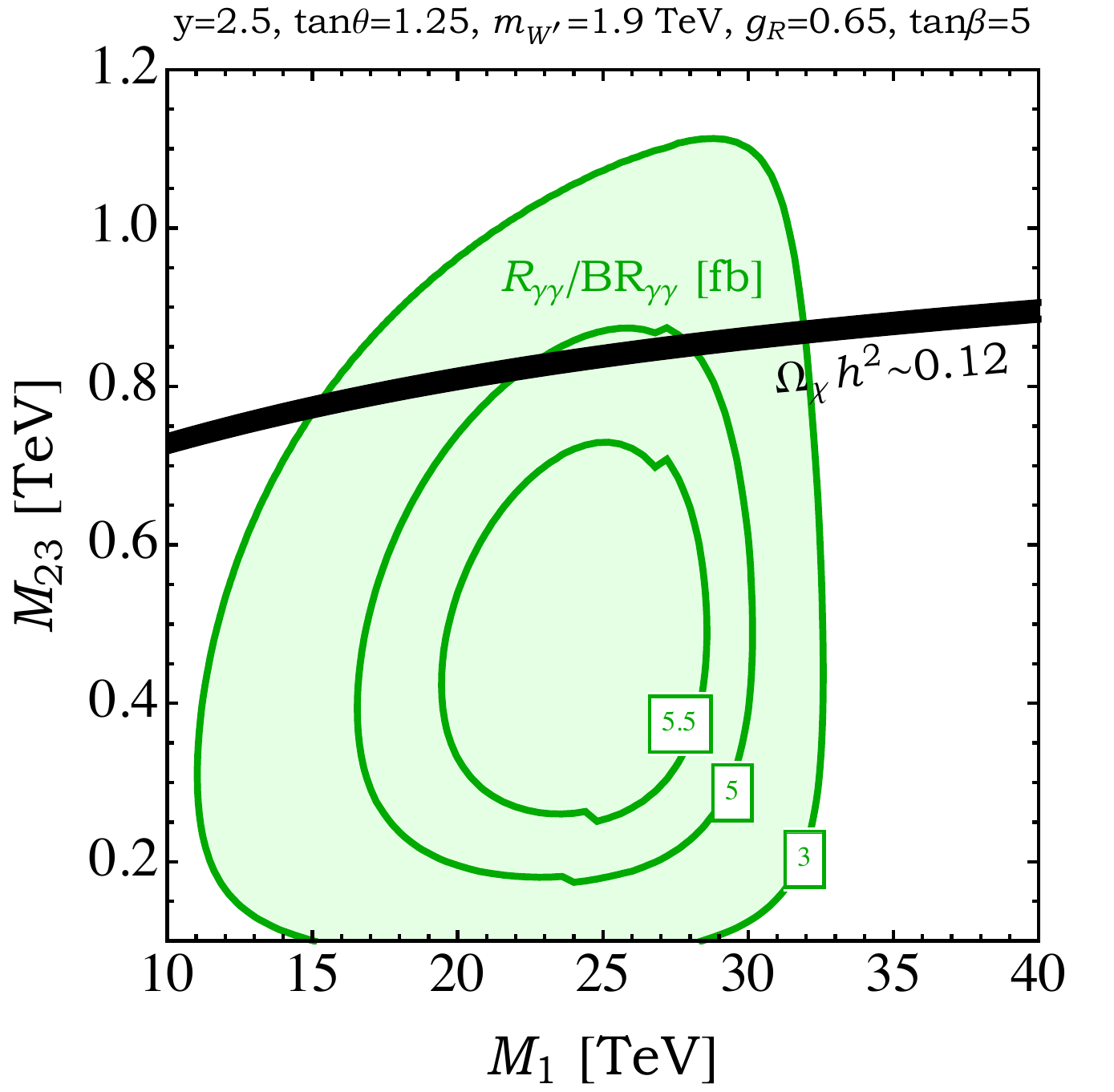}
\includegraphics[width=0.38\textwidth]{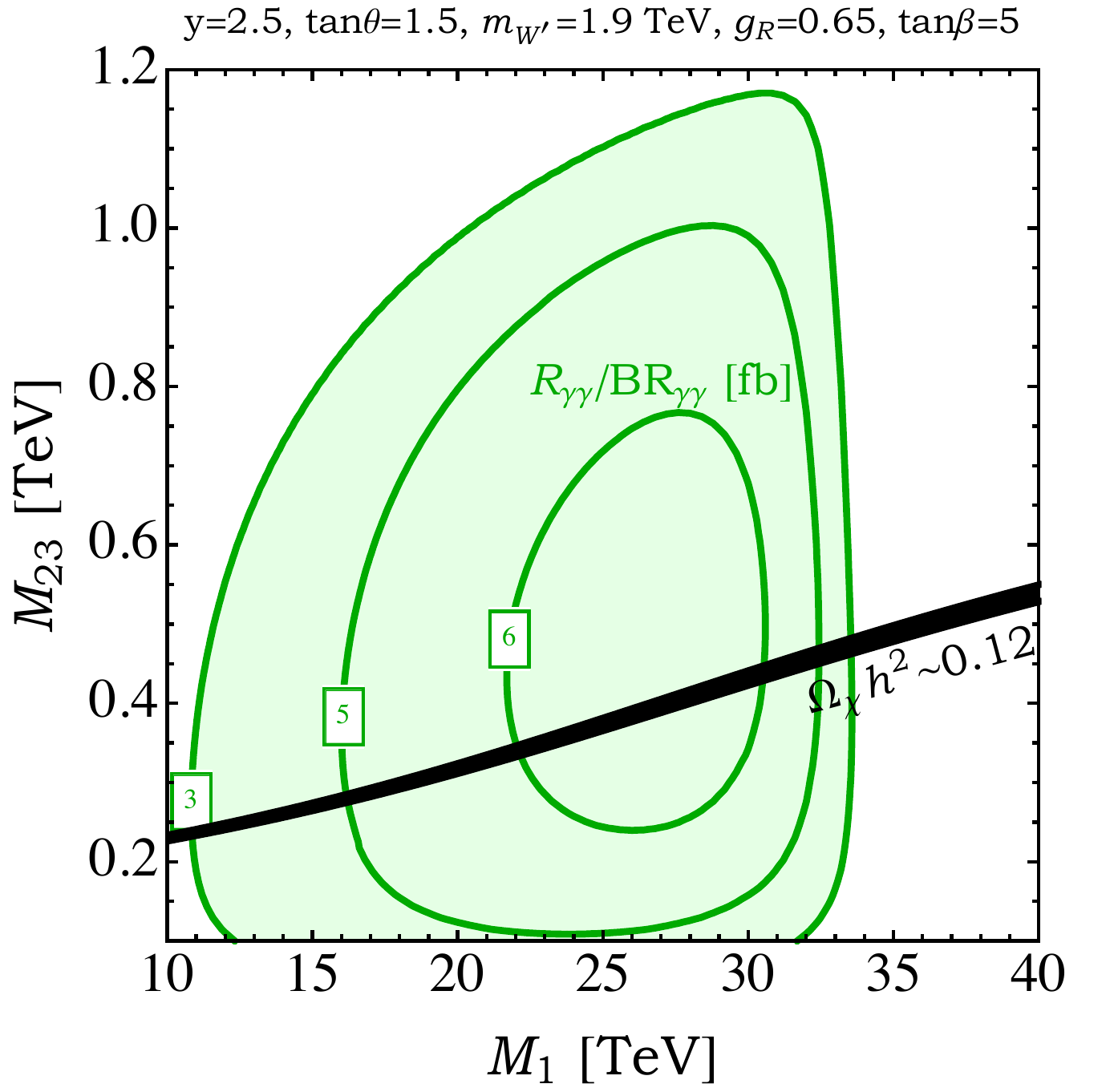} 
\includegraphics[width=0.38\textwidth]{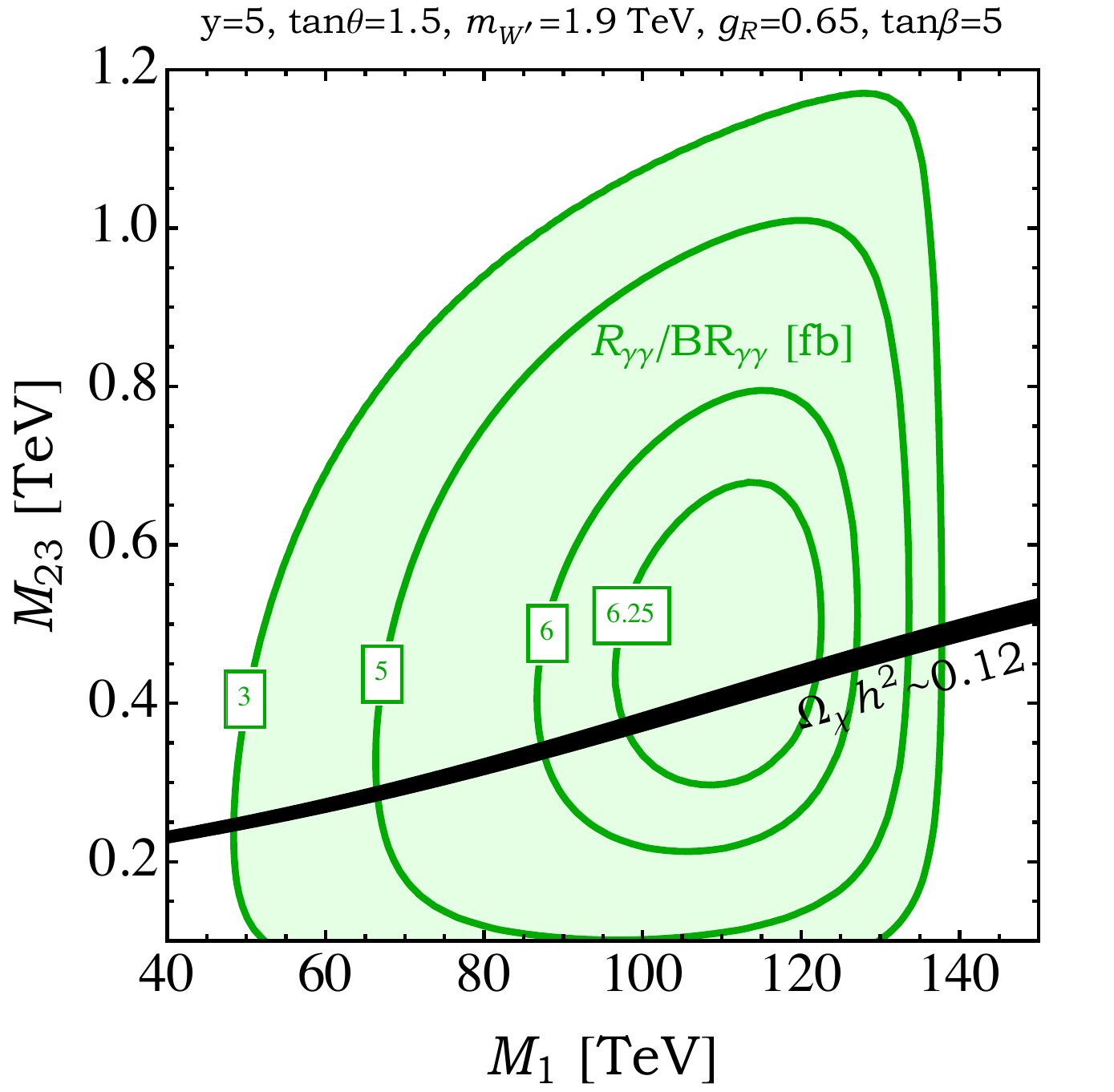} 
\caption{\label{fig:RoverBR} Examples of viable parameter space for the Run 2 diphoton and Run 1 diboson signals. $m_{W^\prime}$, $g_R$, and $\tan{\beta}$ are fixed to values consistent with the diboson excess. Regions favorable with the diphoton excess ($R_{\gamma \gamma} / \text{BR}(\Delta^0 \to \gamma \gamma) \sim $ few fb) are highlighted in green. Also shown is parameter space where the relic abundance of $\x_1$ matches the observed dark matter density (black).}
\end{center}
\end{figure}

The diphoton rate at a center of mass energy $\sqrt{s}$ can be approximated by
\begin{align}
\label{eq:diphotonrate}
R_{\gamma \gamma} &\equiv \sigma (pp \to Z^\prime \to \x_1 ~ \x_{2,3} \to \x_1 ~ \x_1 ~ \Delta^0 (\to \gamma \gamma))
\nl
&= \frac{3 ~ \Gamma_{Z^\prime}}{s ~ m_{Z^\prime}} ~ \text{BR} (\Delta^0 \to \gamma \gamma) ~ \text{BR}(\x_{2,3} \to \x_1 \Delta^0) ~ \text{BR} (Z^\prime \to \x_1 ~ \x_{2,3}) ~ \sum\limits_{q} c_{q \bar{q}} ~ \text{BR} (Z^\prime \to q \bar{q})
~,
\end{align}
where $c_{q \bar{q}}$ are dimensionless partonic integrals involving evaluation of PDFs at a scale $m_{Z^\prime}$~\cite{Franceschini:2015kwy}. We utilize the MSTW2008NLO set of PDFs in evaluating these numerical coefficients~\cite{Martin:2009iq} and require that the rate falls within the range $R_{\gamma \gamma} \sim 1-10$ fb in order to account for the diphoton signal at $\sqrt{s} = 13$ TeV.

Let us first examine the prospect of this production mechanism while remaining agnostic regarding the $\Delta^0$ decay. Fig.~\ref{fig:RoverBR} shows the $\Delta^0$ production rate, $R_{\gamma \gamma} / \text{BR}(\Delta^0 \to \gamma \gamma)$~, for various slices of parameter space, scanning over the bare masses $M_1$ and $M_{23}$ of Eq.~(\ref{eq:DarkLag}). We have switched to the variables $y$ and $\tan{\theta}$ defined through the relations
\be
\lambda_1 \equiv y \sin{\theta} ~,~ \lambda_2 \equiv y \cos{\theta}
~.
\ee
In the top-left panel of Fig.~\ref{fig:RoverBR}, the values of $m_{W^\prime}$, $g_R$, $\tan{\beta}$ (and hence $m_{Z^\prime}$) are fixed within the ranges preferred by the Run 1 diboson excess, while the other panels relax this constraint and consider values of $g_R$ and $\tan{\beta}$ slightly outside the preferred range. Also shown in Fig.~\ref{fig:RoverBR} are regions of parameter space where the relic abundance of $\x_1$ matches the observed dark matter density  $\Omega_\x h^2 \approx 0.1199 \pm 0.0027$~\cite{Ade:2013zuv}. The full set of leading interactions relevant for annihilations and co-annihilations are built in FeynRules~\cite{Alloul:2013bka} and implemented in micrOMEGAs~\cite{Belanger:2001fz}. Proper freeze-out occurs in the black band shown in Fig.~\ref{fig:RoverBR}, and its behavior is mainly governed by the proximity of the lightest neutral state, $\x_1$~, to the singly and doubly charged states, $\x_{1,2}^\pm$ and $\x^{\pm \pm}$~. For sufficiently small mass splittings, annihilations and co-annihilations to SM gauge/Higgs bosons become strong enough to efficiently deplete the thermal $\x_1$ density. 

For each model, we indicate the region optimized for the diphoton signal in green, corresponding to $\Delta^0$ production rates $R_{\gamma \gamma} / \text{BR}(\Delta^0 \to \gamma \gamma) \gtrsim$ 3 fb. Throughout the parameter space shown, the mass splittings $\Delta M_{1,2}$ of Eq.~(\ref{eq:DeltaM}) vary from $\lesssim 100$ GeV to $\gtrsim 1$ TeV, and therefore may introduce significant missing energy in the diphoton channel. The bottom frames of Fig.~\ref{fig:RoverBR} illustrate the effect of varying the value of $\tan{\theta}$ and $y$. Larger values of $\tan{\theta}$ have the potential to increase the $Z^\prime \to \x_1 ~ \x_{2,3}$ branching fraction and hence the overall diphoton rate, but may shift the thermal relic curve away from regions of small $\Delta M_{1,2}$ that are favorable for suppressing extra activity in the diphoton channel. Alternatively, increasing the value of $y$ strengthens the overall signal through the enhancement of $\text{BR}(\x_{2,3} \to \Delta^0 ~ \x_1)$. The diphoton rate, $R_{\gamma \gamma}$~, may account for the ATLAS and CMS excess if $\Delta^0$ possesses a significant branching fraction to photon pairs, $\text{BR}(\Delta^0 \to \gamma \gamma) \gtrsim 1/3$~. Processes relevant for $\Delta^0 \to \gamma \gamma$ are discussed in the following section. 

Loop-induced decays of $\Delta^0$ that are mediated by heavy charged particles with $\order(1)$ couplings naturally result in a diphoton partial width of $\order(1)$ MeV. Therefore, a total $\Delta^0$ width of roughly 45 GeV (as favored by ATLAS) would significantly dilute the branching fraction to photons. Additionally, ATLAS's power to discriminate between a wide and narrow resonance is not statistically significant, and hence, we do not consider the possibility of $\Gamma_{\Delta^0} \sim 45$ GeV further. 

If a large diphoton branching fraction of $\Delta^0$ can be induced, then a consistent explanation of the diphoton and diboson excesses, as well as a standard WIMP cosmology, occurs in the region of parameter space where the thermal relic band is overlayed on the green region favored by the diphoton signal. We will see in the following section that this can easily occur since loops of dark sector charged states tend to dominate the $\Delta^0$ decay. 

Direct detection searches for WIMP dark matter have negligible impact on the models shown. At leading order, WIMP-nucleon scattering proceeds through the tree-level exchange of a $Z^\prime$, which leads to a spin-dependent nucleon scattering rate that is well below the irreducible neutrino background~\cite{Cushman:2013zza}~,
\be
\sigma_\text{SD} \approx \left( \frac{g_{Z^\prime}}{0.1}\right)^2 \left( \frac{3 \text{ TeV}}{m_{Z^\prime}}\right)^4 ~ \left( 6 \times 10^{-45} \right) \text{ cm}^2
~,
\ee
where $g_{Z^\prime}$ is the effective $Z^\prime-\x_1$ coupling.

\subsection{Decay}
\label{subsec:decay}

Under the assumption that mixing with the SM Higgs, $h$~, is negligible, $\Delta^0$ does not couple to the SM fermions at tree-level. Similarly, we will also assume that any trilinear scalar couplings involving both $\Delta^0$ and $h$ are subdominant. Fixing $m_{\Delta^0} \approx 750$ GeV and $m_{\x_1} > 375$~GeV then implies that at tree-level $\Delta^0$ only decays to $W^+ W^-$ pairs, which is suppressed by $m_W / m_{W^\prime}$~. The leading form of this interaction is
\be
\mathcal{L} \supset \frac{\sqrt{2} ~ g_R^3 s_{2 \beta}^2}{g_L^2} ~ \frac{m_W^4}{m_{W^\prime}^3} ~ \Delta^0 ~ W^{+ \mu} W_\mu^- 
~,
\ee
and gives rise to the partial width
\be
\Gamma (\Delta^0 \to W^+ W^-) = \frac{g_R^6 s_{2 \beta}^4}{8 \pi g_L^4} ~ \frac{m_W^4 m_{\Delta^0}^3}{m_{W^\prime}^6} ~ \sqrt{1-4 ~ \frac{m_W^2}{m_{\Delta^0}^2}} ~ \left(1-4~\frac{m_W^2}{m_{\Delta^0}^2} +12 ~ \frac{m_W^4}{m_{\Delta^0}^4} \right)
~.
\ee

We will now consider the radiative decay channels of $\Delta^0$. The Yukawas $\lambda_{1,2}$ of Eq.~(\ref{eq:DarkLag}) allow for interactions between $\Delta^0$  and the neutral and charged states, $\x_{1,2,3}$ and $\x_{1,2}^\pm$~, of Sec.~\ref{sec:model}. Hence, $\x_i$ and $\x_i^\pm$ can mediate loop-induced decays to SM gauge bosons. At leading order, the dominant channels are $\Delta^0 \to \gamma \gamma$, $Z \gamma$, and $ Z Z$ through loops of the singly charged states $\x_{1,2}^\pm$~. Loops of $W$, $W^\prime$, and $Z^\prime$ bosons may also contribute to the width of $\Delta^0$ but are mass suppressed and subdominant for $\order(1)$ values of $\lambda_{1,2}$~. For completeness, we give the expression for the diphoton partial decay width~\cite{Djouadi:2005gi},
\be
\Gamma (\Delta^0 \to \gamma \gamma) = \frac{ \alpha^2 ~ m_{\Delta^0}^3}{256 \pi^3} \Big| \sum\limits_{i=1,2} \frac{\lambda^{(i +)}}{m_i^\pm} A_{1/2} (\tau_i) \Big|^{\, 2}
~,
\ee
where $m_i^\pm$ is the mass of $\x_i^\pm$~, $\tau_i \equiv m_{\Delta^0}^2 / 4 (m_{i}^{\pm})^2$~,
\be
A_{1/2} (\tau_i) \equiv 2 \left[ \tau_i + (\tau_i -1) ~ \text{arcsin}^2 \sqrt{\tau_i} ~ \right] ~ \tau_i^{-2}
~,
\ee
and the $\Delta^0-\chi_i^\pm$ couplings are given by
\be
\lambda^{(i+)} = \frac{1}{2} \left( -\lambda_1 ~ U_{2i} V_{1i} + \lambda_2 ~ U_{1i} V_{2i} \right)
~.
\ee
$\Gamma (\Delta^0 \to Z \gamma)$ and $\Gamma (\Delta^0 \to Z Z)$ take a similar form, aside from minor kinematic factors.

\begin{figure}[t]
\begin{center}
\includegraphics[width=0.38\textwidth]{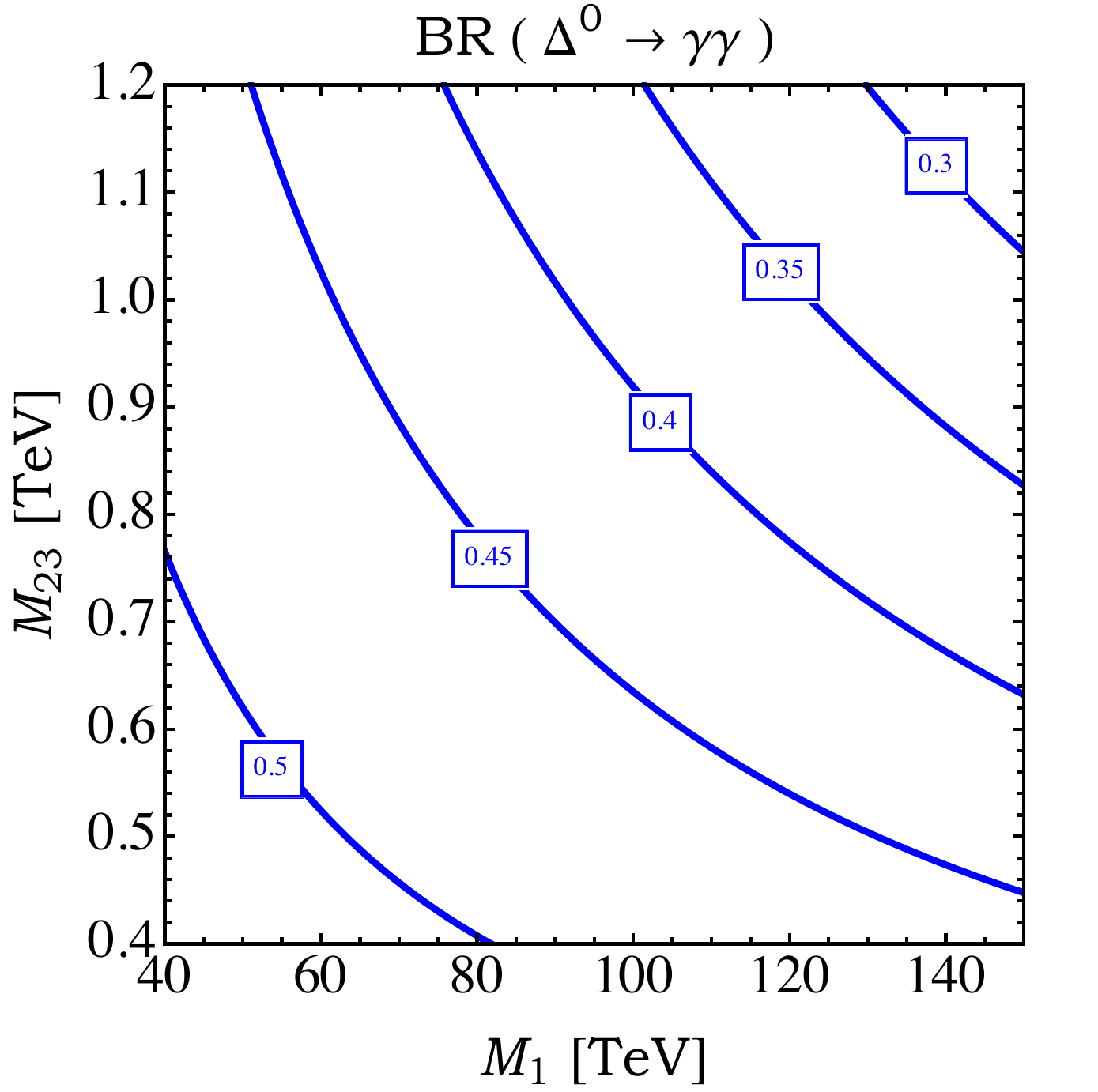} 
\caption{\label{fig:BR} The branching fraction of $\Delta^0 \to \gamma \gamma$ for the set of parameters corresponding to the bottom-right panel of Fig.~\ref{fig:RoverBR}. For $M_{23} \lesssim 400$ GeV, tree-level decays of $\Delta^0$ to dark matter dominate the width. This demonstrates that these models are able to account for the Run 2 diphoton signal for $M_{23} \gtrsim 400$ GeV, across the green regions in Fig.~\ref{fig:RoverBR}.}
\end{center}
\end{figure}

Results for $\text{BR} (\Delta^0 \to \gamma \gamma)$ are shown in Fig.~\ref{fig:BR} for the choice of parameters corresponding to the bottom-right panel of Fig.~\ref{fig:RoverBR}. For $M_{23} \gtrsim 400$ GeV, $m_{\x_1} \gtrsim m_{\Delta^0}/2$~, and the suppression of tree-level contributions to $\Gamma_{\Delta^0}$ allows for a significant branching fraction to $\gamma \gamma$~. This would not be possible if $\Gamma_{\Delta^0} \sim 45$ GeV was enforced. The results are similar for the models corresponding to the other panels of Fig.~\ref{fig:RoverBR}. Therefore, taken together,  Figs.~\ref{fig:RoverBR} and \ref{fig:BR} demonstrate that there are large regions of parameter space for which $R_{\gamma \gamma}$ falls in the required range ($1-10$ fb) for the diphoton signal, while simultaneously accounting for the Run 1 diboson excess and a standard thermal WIMP. Rescaled 8 TeV limits for resonance searches in the $Z \gamma$ final state restrict rates $R_{Z \gamma} \lesssim \order(10)$ fb at $\sqrt{s} = 13$ TeV~\cite{Aad:2014fha}. For our model, $\Gamma (\Delta^0 \to Z \gamma) / \Gamma (\Delta^0 \to \gamma \gamma) \sim 0.6$, and hence there is no tension with these searches.

\section{\textbf{Conclusions}}
\label{sec:conclusions}

Within the framework of a left-right symmetric model, we have presented a UV-complete mechanism to generate both the Run 2 diphoton and Run 1 diboson excesses, while simultaneously allowing for the standard thermal freeze-out of WIMP dark matter. 
The production and decay of a $W^\prime$ boson of mass $\sim 1.8-2$ TeV can account for the excesses observed in Run 1 data through the processes $p p \to W^\prime \to j j \, , \, W Z \, , \, W h$~. The diphoton signal is facilitated by the resonant production and decay of a $3-4$ TeV $Z^\prime$ boson through $pp \to Z^\prime \to \x_1 ~ \x_2 \to \x_1 ~ \x_1 ~ \Delta^0 (\to \gamma \gamma)$~, where $\Delta^0$ is the neutral component of the right-handed Higgs triplet and has a mass of 750 GeV. $\x_{1,2}$ are the lightest and next-to-lightest neutral components of the dark matter multiplets, and interestingly, the charged components of the same multiplets aid in the radiative decay of $\Delta^0$ to SM gauge bosons, providing a large branching fraction to pairs of photons.

In several regions of parameter space, the diphoton signal can be produced at the rate of a few fb in this class of models, sufficient for generating the 750 GeV excess observed in the first data set of Run 2. This rate could be significantly enhanced if one abandons this model as an explanation for the Run 1 diboson signal by lowering both the masses of the $W^\prime$ and $Z^\prime$.

\vspace{0.2in}
\noindent {\em Acknowledgments:} We thank Dan Hooper and Lian-Tao Wang for valuable comments on early versions of this manuscript. AB is supported by the Kavli Institute for Cosmological Physics at the University of Chicago through grant NSF PHY-1125897.

\vspace{-0.03\textwidth}
\bibliography{LRDM}

\end{document}